\newcommand{\ignore}[1]{}
\documentclass[12pt]{article}
\usepackage{cite}
\newcommand\be{\begin{equation}}
\newcommand\ee{\end{equation}}
\newcommand\bea{\begin{eqnarray}}
\newcommand\eea{\end{eqnarray}}
\begin{document}
\def\CN{{\cal N}}
\def\CO{{\cal O}}
\def\CF{{\cal F}}

\begin{titlepage}
\thispagestyle{empty}
\begin{flushright}
BROWN-HET-1335
\end{flushright}
\vskip 1cm
\begin{center}
{\LARGE\bf Hagedorn transition for strings on pp-waves and tori with chemical potentials}
\vskip 1cm
{\large Richard C. Brower$^\dagger$, David A. Lowe$^*$ and Chung-I Tan$^*$}
\vskip .5cm
{\it $^\dagger$Department of Physics \\
Boston University \\
Boston, MA 02215, USA} \\ \vskip .5cm
{\it $^*$Department of Physics \\
Brown  University \\
Providence, RI 02912}\\
\end{center}

\vskip 1cm

\begin{abstract}
  It has been conjectured that string theory in a pp-wave background
  is dual to a sector of $\CN=4$ supersymmetric Yang-Mills theory.
 We study the Hagedorn transition for free strings
  in this background.  We find that the free energy at the transition point
  is finite suggesting a confinement/deconfinement transition in the
  gauge theory. In the limit of vanishing mass parameter the free
  energy matches that of free strings on an 8-torus with
  momentum/winding chemical potential. The entropy in the
  microcanonical ensemble with fixed energy and fixed momentum/winding
  is computed in each case.
\end{abstract}

\end{titlepage}

\section{Introduction}
One of the original motivations for studying string thermodynamics was
to model the confinement/deconfinement phase transition in QCD
\cite{Polyakov:1978vu}. With the advent of a direct correspondence
between supersymmetric gauge theory and string theory in an anti-de
Sitter background \cite{Maldacena:1998re}, this subject has been
re-examined.

As argued in Refs. \cite{Witten:1998qj,Witten:1998zw}, a type of
confinement/deconfinement transition in the strong coupling limit of
large $N$ $SU(N)$ $\CN=4$ supersymmetric Yang-Mills theory on $R^1
\times S^3$ may be mapped to the Hawking-Page phase transition for
black holes \cite{Hawking:1983dh}. True phase transitions only occur
in systems with infinite numbers of degrees of freedom, so it is
necessary to take $N\to \infty$ to see this transition on a compact
$S^3$.  This transition has been further studied in Refs.
\cite{Landsteiner:1999gb,Caldarelli:1999ar,Burgess:1999vb,KalyanaRama:1998cb,Kim:1999sg}.
Unfortunately it is difficult to quantize the string theory directly
in the anti-de Sitter background due to the presence of Ramond-Ramond
flux, so it has not been possible to extend this result to the weak
coupling limit, where one expects to see a Hagedorn-like exponential
increase of the density of states with energy below the deconfinement
transition.

However recently BMN \cite{Berenstein:2002jq} have described how to map a
subsector of the  large $N$ $SU(N)$ $\CN=4$
supersymmetric Yang-Mills theory to a simpler string theory background
where exact quantization is possible, known as the parallel-plane-wave
or pp-wave, with metric characterized by a mass parameter $\mu$,
\be
ds^2 = -2 dx^+ dx^- -   \mu^2 \sum_{i=1}^8 (x^i)^2(dx^+)^2 +
\sum_{i=1}^8 (dx^i)^2 \;.
\label{metric}
\ee
One of the main results of the
present work is to examine free string thermodynamics in this
background, extending earlier results of
Refs. \cite{Greene:2002cd,PandoZayas:2002hh}.

In order to isolate the subsector appropriate in the BMN limit
\cite{Berenstein:2002jq}, it is necessary to introduce, in addition to
the inverse temperature, $\beta$, a chemical potential, $\nu$, for the
longitudinal momentum, corresponding to the large R-charge limit of
the gauge theory. This string background exhibits Hagedorn-like
behavior \cite{Hagedorn:1965st}.  We find that the free energy, as an
analytic function of $\beta$, has a singularity
\be
 F(\beta, \nu; \mu)\sim c
\sqrt{\beta - \beta_H(\nu; \mu)} + {\mbox {regular part}} \;,
\ee
where $c>0$, and the Hagedorn temperature,
$T_H\equiv \beta_H^{-1}$, is a function of both the chemical potential
$\nu$ and the mass parameter. The free energy is finite at the
transition point, suggesting a phase transition rather than a limiting
temperature.

The mass parameter $\mu$ effectively confines oscillations of strings
in the transverse directions $x^i$. In the limit that the curvature
may be neglected (i.e. $\mu\sqrt {\alpha'} \ll 1$) we have found  that all salient features are in fact
already visible in the more familiar context of strings in a toroidal
compactification, contrary to claims in
Refs. \cite{Greene:2002cd,PandoZayas:2002hh}.
We therefore begin by first treating
toroidal compactification with chemical
potentials. This system has been used in the past to model QCD at
finite baryon number chemical
potential in Ref. \cite{Brower:1998an}. Here we obtain a number of results
not previously noted in the literature.

As usual with free string dynamics in the canonical ensemble,
fluctuations may diverge as the Hagedorn temperature is approached
\cite{Mitchell:1987th}, which can invalidate the thermodynamic limit.
In this situation, the microcanonical ensemble is more physically
appropriate. We compute the entropy in
the microcanonical ensemble which clarifies many of our canonical
ensemble results.

\section{Canonical Ensemble and Toroidal Compactification}

The canonical ensemble for free strings has been described in Refs.
\cite{Bowick:1985az,Sundborg:1985uk,Tye:1985jv,Polchinski:1986zf,O'Brien:1987pn,Mclain:1987id,
  Alvarez:1986fw,Alvarez:1987sj,Sathiapalan:1987db,Mitchell:1987th,Atick:1988si,Deo:1989jj,Bowick:1989us,Deo:1989bv}.
We define the multi-string grand canonical ensemble for free strings
in terms of a sum over single string partition functions for bosonic
modes $Z_1^B$ and (spacetime) fermionic modes $Z_1^F$,
\be
\log Z(\beta,\mu,\nu) = \sum_{r=1}^\infty {1\over r} \left( Z_1^B(r \beta,
  r \mu, r \nu) - (-1)^r Z_1^F(r \beta, r \mu, r \nu) \right) \; .
\ee
When we have non-compact spatial dimensions, the behavior near the
Hagedorn temperature will be dominated by the single string $r=1$
term, which we denote $Z_1 = Z_1^B+ Z_1^F$. We define the free energy as
$\beta F = - \log Z \approx -Z_1$. With all dimensions compactified,
as we will see, it is necessary to be more careful in relating the
single string thermodynamics potentials to the full thermodynamic
potentials.

Consider  a system of closed strings in $D$ spatial dimensions, (e.g.,
$D=9$ for the Type II superstring).
Of these, $\bar d$ dimensions are compactified on a
$\bar d$-torus, each with radius
$R_i$, with the remaining $d$ directions uncompactified, i.e., $D=d+\bar d$. The single closed
string free energy density is given by
\be
Z_1(\beta,\lambda,\nu)\!=\frac{\beta V_D}{(2\pi \sqrt {\alpha'})^{D+1}} \int_0^{\infty} \frac{d
\tau_2}{\tau_2^2}
\int_{-1/2}^{1/2} d\tau_1 \;  e^{\displaystyle
 -\frac{\beta^2}{4\pi\alpha'\tau_2}} \;  K(z,\bar z) W(\bar z,  z, \{
R\}, \{\lambda\},
\{\nu\}),
\label{partit}
\ee
where $V_D$ is the spatial volume and $K(z,\bar z)$ is the standard modular
invariant integrand for the uncompactified one-loop partition
function, and we have defined
$z\equiv e^{2\pi i \tau}$ and  $\bar z\equiv e^{-2\pi i\bar\tau}$ with
$\tau = \tau_1 + i \tau_2$.
$W$ is the contribution due to momenta and windings in the compactified
directions, with chemical potentials, $(\lambda_i,\nu_i)$,
\bea
\label{chempot}
&& W(\bar z,  z, \{ R\}, \{\lambda\}, \{\nu\})=\prod_{i=1}^{\bar d}W_0(\bar z,  z,   R_i,  \lambda_i,
\nu_i),\\
&&W_0(\bar z,  z,  R_i, \lambda_i, \nu_i)= \frac{\sqrt{\alpha' \tau_2}}{R_i}
\sum_{p_i,w_i=-\infty}^\infty {\bar z}^{{\alpha'\over
4}\left(\frac{p_i}{  R_i} + \frac{w_i R_i}{\alpha'} \right)^2}{z}^{{\alpha'\over 4}\left(\frac{p_i}{ R_i} - \frac{w_i
R_i}{\alpha'}\right)^2}e^{\displaystyle -\left(\nu_i p_i +\lambda_i
w_i\right)} \; . \nonumber
\eea
In the following we assume the chemical potential terms are
sufficiently small that the sums in these expressions converge.

\subsection{Hagedorn Singularity without Chemical Potentials}

We begin by first setting all chemical potentials to zero and identify
the nature of Hagedorn singularity as the dimension of the
uncompactified directions, $d$, is varied. We shall demonstrate
shortly that the only effect of the chemical potentials will be to
shift the value of the Hagedorn temperature $T_H = \beta_H^{-1}$.  A
Hagedorn singularity occurs as $\beta$ is lowered due to a potential divergence
of the integral in Eq.  (\ref{partit}) at $\tau_2=0$.  To extract the
singularity at the transition temperature, we first
perform the integral over $\tau_1$ to obtain the leading contribution
in the limit $\tau_2 \rightarrow 0$. This is a standard stationary
phase problem.  The relevant factors coming from $K(z,\bar z)
\rightarrow (\sqrt{\tau_2}/|\tau|)^{-D+1} \exp({\beta_H^2 \tau_2 \over
  4\pi \alpha'|\tau|^2})$ and $W \rightarrow
  V_{\bar d}^{-1}  {(2\pi \sqrt {\alpha'})^{\bar d}} (\sqrt{\tau_2}/|\tau|)^{\bar d}$,
lead to for the integrand of $\tau_2$ integral
\bea
&&   \tau_2^{-2}\int_{-1/2}^{1/2} d\tau_1 \; \Big(\frac{|\tau
|}{\sqrt{\tau_2}}\Big)^{d-1}  \exp\left({\beta_H^2 \tau_2 \over 4\pi
\alpha'|\tau|^2}\right)  \;   \nonumber \\
&& \simeq  \tau_2^{(d-3)/2} \int_{-\infty}^{\infty}
{dx} \; (1+ x^2)^{(d-1)/2} \exp\left({\beta_H^2  \over  4 \pi
  \alpha' \tau_2 } \frac{1}{(1+x^2)} \right) \; ,
\label{nzero}
\eea
where in the second expression we have changed the integration variable
to $x =\tau_1/\tau_2$.  The stationary point is $x = 0$ with
Gaussian fluctuations $x^2 = O(\tau_2)$ as $\tau_2 \rightarrow 0$.
Doing the Gaussian integral and inserting the result of (\ref{nzero})
into (\ref{partit}), we get\footnote{Some previous works in the
 literature do not do this integral over $\tau_1$ correctly. For an alternative derivation where the $\tau_1$ integral is done exactly,
see Refs. \cite{O'Brien:1987pn,Deo:1989jj}.}
\be
Z_1 \sim \frac{\beta V_D}{(2\pi \sqrt {\alpha'})^{d}\beta_H V_{\bar d}} \int_0^{\Lambda}
\frac{d\tau_2 }{\tau_2} \tau_2^{d/2} \exp( {\beta_H^2-\beta^2\over
4 \pi \alpha' \tau_2 } )\;,
\label{toneint}
\ee
where  the integral is cut off at the upper limit. The dependence on d, the number of uncompactified spatial dimension,
 is generic while the value for $\beta_H$ depends on specific string
theory in question. For Type II string theory,
$\beta_H = T_H^{-1} = \pi \sqrt{8 \alpha'}$.
As $T\to T_H$ from below,  one
finds, for $D\geq d \geq 0$,\cite{Deo:1989jj,Deo:1989bv}
\be
Z_1 =
\left\{
\begin{array}{ll}
  c_d \; (\beta-\beta_H)^{d/2} +
{\mbox {regular}} \; &: \; d \quad  \mbox{odd} \\
 c_d \; (\beta-\beta_H)^{d/2}\log(\beta-\beta_H)  +
{\mbox {regular}} \; &: \;  d \quad
 \mbox{even}
\end{array}
\right.
\label{freed}
\ee
with $c_d(\beta_H)\sim (-1)^{(d+1)/2}$ and $c_d(\beta_H)\sim
(-1)^{d/2+1}$ for $d$ odd and even respectively. The free energy
always remains bounded for $d\neq 0$, e.g., for $d=1$,
\be
 F(\beta)\sim c
\sqrt{\beta - \beta_H} + {\mbox {regular part}} \;,
\label{freeone}
\ee
where $c>0$.
For the case where all spatial
directions are compactified, $d = 0$, one can show that the coefficient
is exactly $c_0 = -1$, leading to
\be
 Z_1 = - \log(\beta-\beta_H)+
{\mbox {regular}}\; ,
\label{freez}
\ee
and $ Z\simeq const/(\beta-\beta_H)$.  The interpretation for this
case is a little subtle since now the partition function, no longer
dominated by the single string excitations, receives essential contributions
from multi-string states\cite{Brandenberger:1989aj,Deo:1989jj,Deo:1992mp }.

All these potentials are consistent with positive specific heat, $C =
\beta^2(\log Z)''$, as they must for $T<T_H$.  As $T$ approaches $T_H$
from below, both internal energy, $E=-(\log Z)'$, and specific heat
$C$ can diverge for low values of $d$.  In particular, $C/E^2\beta^2$
blows up for $0\leq d\leq 4$ as $T\to T_H$, indicating that energy
fluctuations in the canonical ensemble diverge as the Hagedorn
temperature is approached.  For $d=0$ the free energy $F$ diverges
signaling a limiting temperature. For $d=1$ and $2$, the average
internal energy $E$ diverges. This alone does not signal a limiting
temperature since the free energy $F$ remains finite in these cases,
but is rather a product of the divergent energy fluctuations.  For $d
= 3$ and $4$, the internal energy $E$ is finite, but the specific heat
diverges at the transition point. For $d \ne 0$, the natural
interpretation is in terms of a conventional, most likely second
order, phase transition.  Including interactions may modify this
picture. The Hagedorn transition is associated with a state winding
the imaginary time direction becoming massless
\cite{O'Brien:1987pn,Sathiapalan:1987db, Kogan:1987jd}. As argued in
Ref. \cite{Atick:1988si} interactions can turn this into a first order
transition with a temperature below the Hagedorn temperature.

\subsection{Hagedorn Singularity with Chemical Potentials}

We now proceed to generalize this result to the
situation with nontrivial chemical potentials. For simplicity, we will focus on the case where
chemical potentials are present for only one dimension. First
consider the factor $W_0(\bar z,  z,  R, \lambda, \nu)$. In the limit $\tau_2\to
0$, we can perform the sum over $p$ and $w$ by approximating the sums
by integrals. After rescaling variables as $x=\tau_1/\tau_2$,
$y = \sqrt{\alpha'}\tau_2 (p/R + w R/\alpha')$ and
$z= \sqrt{\alpha'}\tau_2 (p/R - w R/\alpha')$, one finds
\bea
Z_1 &\sim & \int_0^{\Lambda} {d\tau_2 \over \tau_2} \tau_2^{(d-3)/2} \int dx dy
dz \exp\left(
-{\beta^2\over 4 \pi \alpha' \tau_2 } +\right. \nonumber\\
&& \qquad \left. {1\over 2\tau_2 } \left( {4 \pi
 \over 1+x^2} - \pi (y^2+z^2) - i \pi x (y^2-z^2)  - \gamma_+ y
-\gamma_- z \right) \right)
\eea
where
\be
\gamma_+=
(\nu  R + \alpha' \lambda/ R)/\sqrt{\alpha'} \; , \;  \quad  \gamma_-=(\nu  R -
\alpha' \lambda/
R)/\sqrt{\alpha'}  .
\ee
The $x$,$y$ and $z$ integrals are performed first by the  stationary
phase approximation\footnote{The dominate saddle point is  at real
values of  $y = -   (\gamma_+/2 \pi) (1+ix)^{-1}$ and  $z = -
(\gamma_-/2 \pi) (1-ix)^{-1}$ where x lies on the imaginary axis
at $i x = (\sqrt{\beta^2_0 + \alpha' \gamma^2_+} -
  \sqrt{\beta^2_0 + \alpha' \gamma^2_-})/(2 \beta_H)$.}
leading to
\be
Z_1 \sim   \int_0^{\Lambda} \frac{d \tau_2}{\tau_2}
\tau_2^{d/2} \exp \left(-\frac{\beta^2}{4\pi\alpha'\tau_2}
+\frac{1}{16\pi\tau_2}\left(\sqrt {8\pi^2  + \gamma_+^2} +\sqrt{8\pi^2  +
\gamma_-^2}\right)^2 \right) .
\ee
Note that this remains of the same form as (\ref{toneint}), so the nature of
Hagedorn singularity is unchanged. The inverse Hagedorn temperature is
then given by
\be
{\beta_H} =\left(\sqrt {\beta_0^2 + \alpha'  \gamma_+^2}
 +\sqrt{\beta_0^2
+\alpha'
 \gamma_-^2}\right)/2,
\label{hagt}
\ee
where $\beta_0 = \pi \sqrt{8 \alpha'}$. As the chemical
potential $\lambda$ or $\nu$ is
increased, the Hagedorn temperature decreases.\cite{Deo:1989bv}
The general expression for several chemical potentials replaces
$\gamma^2_\pm$ by  $\gamma^2_\pm =  \sum_i (\nu_i  R_i \pm \alpha' \lambda_i/ R_i)^2/\alpha'$.

The free energy is still given by (\ref{freed}) and (\ref{freez}), but now with $\beta_H
= \beta_H(\lambda,\nu)$ given by (\ref{hagt}). The canonical ensemble
remains stable, for $T<T_H=\beta_H^{-1}$.  As in the case without
chemical potential, for $0\leq d\leq 4$, energy fluctuations become large as
the Hagedorn temperature is approached.

Before proceeding to the more interesting situation of the pp-wave
background, it is worth adding several comments.  We first note that
both the location and the nature of the Hagedorn singularity
(\ref{hagt}) can also be understood, as
mentioned earlier,  in terms  of the occurrence of a thermal tachyon
\cite{O'Brien:1987pn,Sathiapalan:1987db,Kogan:1987jd}. We have verified that the
analysis of Ref. \cite{O'Brien:1987pn}
remains valid when chemical potentials
$(\lambda,\nu)$ are introduced.

Consider next the special situation where a nontrivial chemical potential is present only for the momenta, i.e.,
$\nu\neq 0$ and
$\lambda=0$.  Equivalently, one can treat this direction as uncompactified so that the allowed momenta
become continuous.  In this case, due to Lorentz invariance, $Z_1$ can
only depend on the combination $\beta^2 - \tilde \nu^2$, with $\tilde
\nu\equiv \nu R$. It follows that the inverse Hagedorn temperature,
$\beta_H(0,\nu)$, can be directly obtained from $\beta_H(0,0)=\beta_0$
by a Lorentz boost, i.e.,
\be
\beta_H^2(\nu)\equiv
\beta_H^2(0,\nu)=\beta^2_0 + \nu^2 R^2\; .
\label{hagtnu}
\ee
Indeed, this is consistent with Eq. (\ref{hagt}), since
$\gamma_+=\gamma_-= \nu R/\alpha'$ in this limit. Due to T-duality, it
follows that $\beta_H(\lambda,0)$ can also be directly obtained by
such a symmetry consideration. In what follows, we shall treat only
the case where no chemical potentials for winding modes are
introduced, i.e., $\lambda_i=0$.

We also note that conventional chemical potentials in statistical
mechanics are dimensionful, scaled by $\beta$, i.e., $\nu\rightarrow
\beta \bar \nu$. When this is done, one arrives at a Wigner's semi-circle
law for the Hagedorn transition, i.e., when
$\lambda=0$,\cite{Brower:1998an}
\be
\label{eq:semicircular}
T_H^2 +  {\bar\nu}^2= {T_0}^2,
\ee
where $T_0\equiv \beta_0^{-1}= 1/\sqrt{8 \pi^2 \alpha'}$ is the
Hagedorn temperature when chemical potential is absent, and a factor
of $RT_0$ has also been absorbed by $\bar \nu$.  In view of the
previous comment, it is interesting to note that a semi-circle law can
emerge as a consequence of Lorentz invariance involving extra
dimensions.

\section{Canonical Ensemble in the PP-Wave Background}

The pp-wave background and the duality with a sector of
super-Yang-Mills theory has been described in Ref. \cite{Berenstein:2002jq}
and references therein. The canonical ensemble with chemical potential
for the longitudinal momentum was described in Ref. \cite{Greene:2002cd}, and without chemical potential in Ref. \cite{ PandoZayas:2002hh,Sugawara:2002rs},
\be
Z_1(a,b;\mu) = {\rm Tr}_{\cal H} e^{-a p_+ - b p_-}\;,
\ee
where  $p_+ = p^- = (p^0-p^9)/\sqrt{2}$ and
$p_- = p^+ = (p^0+p^9)/\sqrt{2}$.
Hence we should identify
\be
\beta  = \frac{b+a}{\sqrt{2}}  \;\; , \;\; R\nu  = \frac{b-a}{\sqrt{2}}\;,
\ee
in order  to compare with the results of the previous
section. We can of course scale $R\to \infty$ when the longitudinal
direction is non-compact,
keeping $\tilde \nu = R \nu$  and $p^9= Q/R$ fixed. In what follows, we shall simply replace $R\nu$ by $\nu$ and $p^9$ by $Q$, so that
$\nu$ and $Q$ are now dimensionful.

The lightcone Hamiltonian is given by
\be
p_+=  \frac{1}{ \alpha' p_- } \left( \omega_0(m)
  \left(N_0^B+N_0^F\right)  + \sum^\infty_{n=1}  \omega_n(m)
\left (N_n^B+N_n^F+
\tilde N_n^B+\tilde N_n^F \right) \right) ,
\ee
where
\be
\omega_n(m) = \sqrt {n^2 + m^2},
\ee
with mass $m=2\alpha' \mu p^+$. It follows that
\be
Z_1(a,b;\mu)\propto   \sum_{N_n, \cdots, \tilde N_n} \int_0^{\infty}
d p_-  e^{-b p_- - a p_+}
\ee
subject to the level-matching constraint
\be
\sum_n n (N_n-\tilde N_n)=0.
\ee
So far we are treating the longitudinal direction as non-compact.
However, we can allow for winding in this direction as well if we so
choose. The relevant string compactifications have been described in
Ref.  \cite{Bertolini:2002nr,Mukhi:2002ck}.

As shown in Ref. \cite{Greene:2002cd}, we may then change variables to
$\tau_2 = a/2\pi \alpha' p_-$ and introduce an auxiliary integration
variable $\tau_1$ to implement the level-matching constraint to give
\be
Z_1(a,b;\mu) = {a V_L\over 4\pi^2  \alpha'} \int_0^\infty\frac{d\tau_2}{\tau_2^2}
\int_{-1/2}^{1/2} d\tau_1  e^{-\frac{ab}{2\pi
\alpha' \tau_2}}    \left( { \Theta_{1/2,0}
 (\tau,\bar \tau, \mu a /2\pi \tau_2) \over \Theta_{0,0}
 (\tau,\bar \tau, \mu a /2\pi \tau_2) }\right)^4
\ee
where  the generalized theta functions are defined in
\cite{Takayanagi:2002pi},
$\tau=\tau_1+i\tau_2$, and $V_L$ is the longitudinal volume.
The high temperature behavior may be read off by looking at the $\tau_2 \to 0$
limit. It is convenient to perform a modular transformation on the
theta function as in Ref. \cite{PandoZayas:2002hh} to analyze this limit.
The integrand becomes
\be
{1\over \tau_2^2} \exp\left(- \frac{ab}{2 \pi \alpha' \tau_2} \right)
\exp\left( \frac{16 \mu a \tau_2}{4\pi |\tau|^2}  \left( f(\frac{\mu
 a|\tau|}{2\pi\tau_2},0) - f(\frac{\mu
  a|\tau|}{2\pi\tau_2},\frac{1}{2}) \right) \right)
\ee
where
\be
f(x,\alpha) \equiv 2 \sum_{l=1}^\infty \frac{e^{2 \pi i l \alpha}}{l} K_1(2
\pi l x )
\ee
with $K_1$ a modified Bessel function of the second kind
\cite{PandoZayas:2002hh,Greene:2002cd}.

Performing the $\tau_1$ integral as in (\ref{nzero}), we are left
with
\be
Z_1(a,b;\mu) \sim \int^{\Lambda}_0 \frac{d\tau_2}{\tau_2} {\tau_2^{1/2}}
\exp\left({\beta_c^2-2ab \over 4 \pi \alpha' \tau_2 } \right)
\label{pppart}
\ee
where $\beta_c^2/\alpha'$ depends only on $\mu a$,
\be
{\beta_c^2(\mu a) } = 16\alpha' \mu a   \left( f(\frac{\mu
  a}{2\pi},0) - f(\frac{\mu
  a}{2\pi},\frac{1}{2}) \right)\;.
\label{pphag}
\ee

As the product $ab$ is  increased,  contribution from  $\tau_2\to 0$ leads to a singularity for the
free energy
\be
F \sim \left(2 ab -\beta^2_c(\mu a)\right)^{1/2} + {\mbox {regular}},
\label{ppfree}
\ee
the same basic form as on the torus with $d=1$, (\ref{freeone}).
Since the free energy is finite at the transition point, there may be
a true phase transition at this temperature.

Our result (\ref{ppfree}) differs from that of Refs.
\cite{PandoZayas:2002hh,Greene:2002cd} due to our improved treatment
of the $\tau_1$ integral which yields a different power of $\tau_2$ in
the prefactor in (\ref{pppart}). The expression for the Hagedorn
temperature is in complete agreement with that given in Refs.
\cite{PandoZayas:2002hh,Greene:2002cd}.  Those papers also define the
Hagedorn temperature as $\beta_c^{-1}$, which corresponds to the
proper temperature at which the partition function is singular. In the
following we will stick with our definition of Hagedorn temperature as
the critical value of $\beta^{-1}$.

The inverse Hagedorn temperature   $\beta_H(\nu;\mu)$ is specified by the solution to
\be
 \beta_H^2 -  \nu^2
= \beta_c^2\Big( \frac{\mu ( \beta_H -  \nu)}{\sqrt{2}} \Big)\; .
\label{hagPP}
\ee
To gain a qualitative understanding for the solution, let us
consider two simple limits:\footnote{We are interested in the region where
  $\nu<0$, corresponding to having positive $Q>0$.   }
$ |\nu| \ll \mu^{-1}$ and $|\nu| \gg \mu^{-1}$.  Here we simply address the behavior of $\beta_H(\nu;\mu)$
while leaving the interpretation to the next section where we discuss the microcanonical ensemble.

For the first limit, we need $\mu a \ll 1$ when (\ref{pphag}) becomes
\be
\beta_c^2(\mu a) \simeq \beta_c^2(0)=8 \pi^2 \alpha'=\beta_0^2\;.
\label{ppshag}
\ee
Thus as we approach the flat space limit~\footnote{In fact in this
  limit, there is an accumulation of sub-leading Hagedorn-like
  singularities, with spacing $\sim \mu\alpha'$, causing a non-uniformity in the limits $T \rightarrow
  T_H$ and $\mu \rightarrow 0$. In this paper, we consider only the ``near flat space limit" where $\mu\sqrt\alpha'<<1$ with $\mu\neq 0$.
In the strict $\mu = 0$ case
at fixed
  $T < T_H$ the free energy will exhibit the singularity for zero
  compact dimensions,  i.e., $d=9$.}, we reproduce the same Hagedorn
temperature as obtained for  torus compactification,
\be
\beta_H^2(\nu;\mu)\simeq \beta_0^2 + \nu^2\; ,
\ee
which led to the semi-circular law (\ref{eq:semicircular}).
As in the torus case, the thermodynamic ensemble is stable in this limit.

In the second limit, one needs  $\mu a \gg 1$, and
(\ref{pphag}) becomes
\be
\beta_c^2(\mu a) \simeq  32 \sqrt{2 \pi \mu a} \alpha' e^{-\mu a/\pi}.
\label{ppbhag}
\ee
For $\nu<0$ and $|\nu|$ large, this leads to
\be
\beta_H^2(\nu;\mu)\simeq \nu^2 + 4 \beta_0^2\sqrt{\sigma^3\mu|\nu|} e^{-\sigma\mu |\nu|},
\label{ppchag}
\ee
where  $\sigma=\sqrt 2  /\pi$.  In the limit $\mu|\nu|\rightarrow \infty$, one has
$\beta_H(\nu;\mu)\simeq  |\nu| + 0(e^{-\sigma\mu |\nu|})\;.$
In particular, the Hagedorn temperature
$\beta_H^{-1}$ decreases as
$|\nu|$ increases.  We have checked the  eigenvalues of the matrix of second derivatives of
(\ref{ppfree}) are all negative, as required for stability.

If instead of treating the longitudinal direction as noncompact, we
had compactified and allowed for spatial windings, the prefactor
$1/\tau_2^{1/2}$ in (\ref{pppart}) would change to $1/\tau_2$, and
(\ref{ppfree}) should be changed to
\be
\beta F \sim  \log\left(2 ab -\beta^2_c(\mu a)\right).
\label{ppcfree}
\ee
The same conclusions hold regarding the stability of the ensemble. Now
the free energy diverges at the transition point, so in this case the
Hagedorn temperature could be interpreted as a limiting temperature for
the ensemble. However once interactions are included we expect to see
a first order phase transition as mentioned above, and this is
naturally associated with a confinement/deconfinement phase transition.

\subsection{Yang-Mills Dual}

Finally let us discuss the implications of the results for the dual
Yang-Mills theory. BMN \cite{Berenstein:2002jq} map
string theory $p_+$ and $p_-$ to Yang-Mills energy $ E_{YM}$
and R-charge $Q_{YM}$
as
\be
\frac{\sqrt{2} p^-}{\mu} = E_{YM}-Q_{YM} \quad \quad \sqrt{2}\mu  p^+=
\frac{E_{YM}+Q_{YM}}{R_{AdS}^2 },
\label{symmap}
\ee
where one has in mind the large $N$ limit with $Q_{YM}^2/N$,
$E_{YM}-Q_{YM}$,  and
the Yang-Mills coupling $g_{YM}$  held fixed. Here
the radius of curvature of the $AdS^5$ space is $R_{AdS}^4 = 4\pi g_{YM}^2 N
{\alpha'}^2$, which goes to infinity in this limit, so that $p^+$ and
$p^-$ remain fixed. Note that $\mu$
may be scaled to $1$ using the longitudinal boost symmetry $p^- \to \alpha p^-$,
$\mu \to \mu/\alpha$, $p^+ \to \alpha p^+$ of the metric
(\ref{metric}) and does not appear
directly in the Yang-Mills theory.

In the Yang-Mills theory $E_{YM}$ corresponds to the dimensionless
energy on the space $S^3 \times R$ in units of the radius, $R_{S^3}$,
and $Q_{YM}$ corresponds to the R-charge.  We define the
Yang-Mills temperature to be conjugate to $E_{YM}$ and the chemical
potential to be conjugate to $Q_{YM}$.  With the identifications
(\ref{symmap}) we find
\be
\beta_{YM} =  \frac{\mu a}{\sqrt{2}}+ \frac{b}{\sqrt{2}R_{AdS}^2\mu},
\quad\quad
\nu_{YM} =  \frac{b}{\sqrt{2}R_{AdS}^2\mu}-\frac{\mu a}{\sqrt{2}},
\label{ymvar}
\ee

We wish to interpret the Hagedorn transition as a kind of
confinement/deconfinement transition for the large $N$ gauge theory on a
compact space. Since we have a conformal field theory, the scale is
set by the size of the $S^3$ rather than a dynamically generated scale
as in real QCD. Note that both $\beta_{YM}$ and $\nu_{YM}$ are dimensionless, with
$\beta_{YM}+\nu_{YM}= (\beta+\nu)/\mu R^2_{AdS}$ and $\beta_{YM}-\nu_{YM}=\mu (\beta-\nu)$.
Of course the invariant properties of the canonical ensemble are
unaffected by this redefinition of parameters. For example
we have
\be
(\beta^2_{YM} - \nu^2_{YM}) =  (\beta^2 - \nu^2)/R_{AdS}^2.
\ee
In the near flat space
limit, a semi-circle law   still holds for the Yang-Mills parameters at the Hagedorn singularity,
\be
T^2_{YM} + \bar \nu_{YM}^2 = T^2_{0;YM} \;,
\ee
where the radius of the circle is set by $T_{0;YM}=1/ \sqrt{8 \pi^2
  \alpha'_{YM}}$ with  $\alpha_{YM}^{'}=  R_{S^3}^2/\sqrt{g_{YM}^2 N}$, the
inverse Yang-Mills  string tension relative to the size of
$S^3$, and  $\bar \nu_{YM}$, the chemical potential after an appropriate re-scaling, $\nu_{YM}\to \beta_{YM} \bar\nu_{YM}/T_{0;YM} $.
Beyond the Hagedorn transition the perturbative string theory picture seems to break down at fixed small string coupling, and it
is natural to conjecture the proper definition of the string theory is in
terms of the large $N$ gauge theory in the deconfining phase. This is
reminiscent of the type  of transition envisaged in Ref. \cite{Atick:1988si}
but the details are rather different.

In a strict pp-wave limit, the inverse Yang-Mills string tension
$\alpha_{YM}^{'}$ vanishes, so the Hagedorn behavior is better
described in term of $a$ and $b$, which are fixed. In this limit, $\mu
a/\sqrt 2\simeq \beta_{YM} \simeq - \nu_{YM}$ and $\sqrt 2
b/R^2_{AdS}\mu=\beta_{YM}+\nu_{YM} \to 0$. It is useful to treat $a$
as a new inverse temperature, conjugate to $E_{YM}-Q_{YM}$, and $b$ as
a new chemical potential, conjugate to $E_{YM}+Q_{YM}$.  In this new
language, as the temperature vanishes, $a\to \infty$, one approaches
the ``ground state" where $E_{YM}-Q_{YM}=0$.

In the limit $b$ large (large
chemical potential), the Hagedorn transition  now corresponds to the near flat space limit (\ref{ppshag}) with
\be
a \approx \frac{\beta_0^2}{2 b},
\ee
where the new transition temperature $a^{-1}$ increases
without bound.  In the small $b$ limit (small chemical potential), the transition is now
the same as $\mu a >> 1$, (\ref{ppbhag}), with
\be
\mu a \sim - \pi \log( b/\mu \alpha')\;,
\ee
so that the transition temperature decreases as $b$ decreases.

It is also interesting to note the free energy given by Eq.
(\ref{ppcfree}) matches exactly with that calculated in a simplified
model for the large $N$ $\CN=4$ supersymmetric Yang-Mills theory on
$S^3$ \cite{Sundborg:1999ue}. (See also Ref.  \cite{Polyakov:2001af}
for closely related results.)  The results of Ref. \cite{Lowe:1995nm}
show that when weak string interactions are included with non-compact
spatial dimensions the density of states shifts to that of the fully
compactified case.  This supports the comparison of the result of Ref.
\cite{Sundborg:1999ue} to Eq. (\ref{ppcfree}) rather than Eq.
(\ref{ppfree}). One might then wonder if the Hagedorn temperature
should then be interpreted as an unattainable limiting temperature or
whether a real phase transition happens.  In Ref.
\cite{Sundborg:1999ue} it was found for the model considered there
that finite $N$ smooths out the singularity in the free energy. A
cross-over to a free energy proportional to $N^2$ occurs above the
transition point, yielding an interpretation in terms of a
confinement/deconfinement transition.  As we have mentioned in the
case at hand, string interactions also open up the possibility of a
first order phase transition at a temperature just below the Hagedorn
temperature which cuts off the limiting temperature behavior, and a
very similar physical picture may emerge.

To describe the physics above the transition point we would need to
generalize the Hawking-Page transition to the BMN limit. In fact, a
pp-wave solution as a Penrose limit of the AdS Schwarzschild
background has been constructed in Ref. \cite{PandoZayas:2002rx}.
However a pp-wave background
cannot give rise to a background with a horizon \cite{Hubeny:2002pj}.
Instead the background is
singular, and difficult to quantize, and hence it is unclear
how a
gravitational entropy of order $N^2$ arises.

With the longitudinal direction of the string theory compactified, one
should still have duality with the large $N$ gauge theory described in Ref.
\cite{Bertolini:2002nr,Mukhi:2002ck}, and the free energy up to the Hagedorn point
should be described by the general form given by Eq. (\ref{ppcfree}). As we have
seen the nature of the singularity is dependent only on the number of
non-compact directions, and only the Hagedorn temperature itself is
sensitive to the details of the compactification.  The
free energy now diverges at the transition, and the same issues as
already discussed above arise.

\section{Microcanonical Ensemble}

In the previous section, we studied the canonical ensemble and found
that there are
large energy fluctuations as the system approached the Hagedorn temperature,
implying the system did not have a smooth thermodynamic limit for a
fixed temperature. These difficulties may be circumvented by studying
the system in the more fundamental microcanonical ensemble. The non-smoothness of the
thermodynamic limit then shows up in the guise of long-string
ensembles that dominate the entropy
\cite{Deo:1989bv,Deo:1992mp,Lowe:1995nm}. In particular, a better
understanding can be achieved in this approach in terms of the
behavior of the entropy as $T_H\rightarrow 0$.

Using the results for the canonical ensemble, we can apply an inverse
Laplace transform to obtain the entropy in the microcanonical ensemble
for fixed $E$ and fixed charge/angular momentum $Q$,
\be
\Omega(E,Q)=\int_{-i\infty}^{i\infty} \frac{d\nu}{2\pi
 i}\int_{-i\infty}^{i\infty} \frac{d\beta}{2\pi i} e^{\beta E+\nu Q}
Z(\beta,\nu) \; ,
\label{omdef}
\ee
where the contour in $\beta$ is  taken to lie to the right of any singularity
in $Z$ while the contour in $\nu$ is along the imaginary axis.
 In the following we will denote the integrand obtained after
performing the $\beta$ integral but before performing the $\nu$
integral as $Z(E,\nu) e^{\nu Q}$, where $Z(E,\nu)$ may be regarded as the partition
function of an ensemble with fixed total energy, but in contact with a
charge reservoir with fixed chemical potential $\nu$.

\subsection{Torus Compactification}

We will restrict our attention to the case of a single chemical
potential $\nu$. For $d>0$, at large $E$
the integral (\ref{omdef}) receives its leading
contribution from the nearest singularity(\ref{freed}), in $Z_1$
leading to
\be
Z(E,\nu) \sim \frac{e^{\beta_H(\nu) E}}{E^{d/2+1}} \; ,
\ee
where $\beta_H(\nu)$ is given by (\ref{hagtnu}).  For $d=0$,  since the coefficient in front of $\log
(\beta-\beta_H)$ in Eq. (\ref{freez}) is exactly $-1$,  after exponentiation giving $Z(\beta,\nu) =
\mbox{const}/(\beta -\beta_H(\nu))$, we obtain
\be
Z(E,\nu) \sim e^{\beta_H(\nu) E}\; .
\ee

Since we are discussing an ensemble in contact with a charge reservoir,
it is necessary to check the ensemble is stable with respect to
fluctuations. In the case at hand this reduces to the condition that
$d^2 \beta_H(\nu) / d \nu^2 > 0$. It is straightforward to check this
is satisfied for all values of $\nu$. This implies the integral
(\ref{omdef}) will be dominated by the stationary point where
\be
Q =- \frac{\partial \beta_H(\nu) }{\partial \nu} E,\quad{\mbox{or}}\quad \nu
=-{\beta_0} \frac{Q}{\sqrt{E^2-Q^2}}\;,
\label{saddle}
\ee
where, as noted earlier, we have absorbed the dependence on $R$ by replacing  $Q/R$
by  $Q$ and $\nu/R$ by $\nu $.  Ignoring prefactors that are power-like in $E$ and $Q$ and $d$
dependent,
we find
\be
\Omega(E,Q)  \sim e^{\beta_0 \sqrt{E^2 -  Q^2}}\;.
\label{flatmicro}
\ee
As $Q \to E $, the linear growth in the entropy $S=\log \Omega$ with
energy disappears. The temperature in microcanonical ensemble, $T= (\partial S/\partial E)^{-1}$, which at this level of
approximation is simply the Hagedorn temperature, becomes
\be
T_H \simeq
\beta_0^{-1}\sqrt{1-(Q/E)^2},
\ee
and it  goes to zero as
$Q$ approaches its maximal value.
As noted in Ref. \cite{Brower:1998an} this is a useful model for the finite
temperature and baryon number chemical potential  phase diagram in QCD.

\subsection{PP-Wave Background}

Precisely the same arguments carry over to the microcanonical ensemble
in the pp-wave background, with the replacement of $\beta_H(\nu)$ by
$\beta_H(\nu;\mu)$ given by the solution of
(\ref{hagPP}). Expressing this in terms of the thermodynamic potential
$Z(E,\nu)$ we find
\be
Z(E,\nu) \sim
\left\{
\begin{array}{lll}
 \exp\left( \beta_H(\nu;\mu) E \right)/E^{3/2}  \; &: \quad
\mbox{non-compact  longitudinal  direction} \\
\exp\left( \beta_H(\nu;\mu) E \right) \; &: \qquad \mbox{ compact
 longitudinal  direction}
\end{array}
\right .
\label{ppnumicro}
\ee

In the limit $\mu a\ll 1$, we recover the standard flat space
results for the entropy, obtained by substituting $ \beta_H = \sqrt{
\beta_0^2 + \nu^2}$ into (\ref{ppnumicro}) and integrating over
$\nu$ in (\ref{omdef}). Therefore the entropy can be obtained  simply from
(\ref{flatmicro}), subject to the condition
\be
\mu \sqrt{\alpha'}\ll \sqrt{(E-Q)/(E+Q)}.
\ee
This indicates the dual sector of the Yang-Mills
theory should show an exponentially rising density of states with
energy. The
scale of the inverse string tension $\alpha'$ in string variables translates
into inverse string tension $\alpha_{YM}^{'}=(\sqrt{g_{YM}^2 N})^{-1} R_{S^3}^2$ in Yang-Mills
variables (\ref{symmap}), where $R_{S^3}$ is the radius of the $S^3$, with Hagedorn temperature becoming
\be
T_H \simeq
(8\sqrt{\pi\alpha'_{YM}} )^{-1}\sqrt {1-(Q_{YM}/E_{YM})^2}.
\ee
For finite $N$ we expect to see a  cutoff on
the range of energy exhibiting this behavior, as suggested by the toy
model of Ref. \cite{Sundborg:1999ue}, and already discussed above.

When $\mu a\gg 1$, $\beta_H(\nu;\mu)$ is given by (\ref{ppchag}). We
may then compute (\ref{omdef}) in a saddle point approximation where
\be
\mu a\sim  \pi \log (\frac {E+Q}{E- Q}) +\pi \log(\mu^2\alpha').
\ee
Ignoring
prefactors that are powerlike in $E$ and $Q$, one arrives at
\be
\Omega(E,Q) \sim \exp\left( \frac{\pi}{\mu} (E-Q) ( \log
\frac{E+Q}{E-Q} +\log(\mu^2\alpha') )
 \right) \;,
\label{ppmicro}
\ee
subject to the condition
\be
\mu \sqrt{\alpha'}\gg \sqrt{(E-Q)/(E+Q)}     \;.
\ee
 This limit corresponds to $(E-Q)/(E+Q)\to 0$.
As $Q\to E$,  the  entropy decreases
as $({\pi}/{\mu} )(E-Q)\log (E-Q)$,
and the Hagedorn temperature decreases as
\be
T_H\sim
(\mu /\pi)(\log({E-Q}))^{-1}\:.
\ee
It is interesting to note the pp-wave mass
parameter $\mu$ now sets the scale of the Hagedorn-like growth in the
number of states. When we translate this into Yang-Mills variables
(\ref{symmap}) we obtain an inverse string tension proportional to
$R_{S^3}^2$ only.

We note in passing that when the subleading corrections to the saddle
point approximation are computed, we find the temperature can become
negative in the microcanonical ensemble. Likewise the specific heat is
not well-behaved. This is symptomatic of the long string states that
dominate the microcanonical ensemble at high energies
\cite{Deo:1989bv,Deo:1992mp,Lowe:1995nm}, and is a sign of large
fluctuations in
the canonical ensemble, rather than some inconsistency in the
microcanonical ensemble.

\section{Conclusion}

We have studied the thermodynamics of IIB free strings as one
approaches the Hagedorn transition from the low temperature phase in
both the pp-wave background and in flat space on compact tori with
non-zero chemical potential. A comparison reveals that the pp-wave
transition is well characterized by the flat space example on an
8-torus with a chemical potential in the longitudinal momentum playing
the role of the R charge. Apparently the harmonic ``potential'' in the
transverse co-ordinates, $\mu\sum x^i x^i$,  of the pp-wave background
effectively acts like a compact space. This correspondence for the
Hagedorn temperature and the free energy become an identity if we take
the near flat space limit $\mu \sqrt {\alpha'}\ll 1$.

In both cases the free energy exhibits a  singularity, $F
\sim c \sqrt{\beta - \beta_H} + \mbox{regular}$, $c>0$, consistent with the onset of a phase
transition rather than a limiting temperature.  This supports the
interpretation of a Hagedorn transition as a close analogue of the
conventional picture for the confinement/deconfinement transition.  Indeed the real
interest in this calculation for the pp-wave is the existence of a
dual ${\cal{N}} = 4$ super Yang-Mills field theory on $R \times S^3$.
On the Yang-Mills side it is possible to go to the high temperature
side of the Hagedorn transition where the entropy is expected to rise
sharply from $O(1)$ to $O(N^2)$. At $N = \infty$, the $O(1)$ term
which exhibits the Hagedorn singularity does not have to be connected
analytically to the $N^2$ term.  However at finite N (when strings begin
to interact) the two regions should be part of a single thermodynamics
description.

Understanding this transition on the string side is a long standing
puzzle in interacting string theory. A natural supposition as
suggested by Witten~\cite{Witten:1998qj,Witten:1998zw} is that the
Hawking-Page phase transition for black holes that occurs in the
strong coupling description of strings in $AdS^5$ causes interacting
strings to deconfine. A recent paper by Zayas and Sonnenschein
\cite{PandoZayas:2002rx} has taken the pp-wave limit of the black hole
metric for the $AdS^5$ space.  A difficult but interesting problem is
to explore this trans-Hagedorn phase as strings propagate in this
background.  In view of the exact mapping for gravity/gauge duality in
the pp-wave background, this appears to be a good setting in which to
further explore these phenomena in string dynamics at extreme
temperature.  Ultimately one goal is to shed some light on QCD at
finite temperature and chemical potential
\cite{Alford:2002ng,Rajagopal:2000wf}, a region of phase space out of
reach of present non-perturbative (i.e. lattice) methods but of
growing experimental and phenomenological interest in heavy ion
collisions and ultra dense astronomical objects.

\bigskip \centerline{\bf Acknowledgements}
\noindent
This research is supported in part by DOE
grants DE-FE0291ER40688-Task A and DE-FG02-91ER40676.

\bibliography{hagbib}
\bibliographystyle{hunsrt}

\end{document}